# Image Segmentation of Zona-Ablated Human Blastocysts

Md Yousuf Harun, M Arifur Rahman, Joshua Mellinger, Willy Chang, Thomas Huang, Brienne Walker, Kristen Hori, and Aaron T. Ohta

*Abstract*— Automating human preimplantation embryo grading offers the potential for higher success rates with in vitro fertilization (IVF) by providing new quantitative and objective measures of embryo quality. Current IVF procedures typically use only qualitative manual grading, which is limited in the identification of genetically abnormal embryos. The automatic quantitative assessment of blastocyst expansion can potentially improve sustained pregnancy rates and reduce health risks from abnormal pregnancies through a more accurate identification of genetic abnormality. The expansion rate of a blastocyst is an important morphological feature to determine the quality of a developing embryo. In this work, a deep learning based human blastocyst image segmentation method is presented, with the goal of facilitating the challenging task of segmenting irregularly shaped blastocysts. The type of blastocysts evaluated here has undergone laser ablation of the zona pellucida, which is required prior to trophectoderm biopsy. This complicates the manual measurements of the expanded blastocyst's size, which shows a correlation with genetic abnormalities. The experimental results on the test set demonstrate segmentation greatly improves the accuracy of expansion measurements, resulting in up to 99.4% accuracy, 98.1% precision, 98.8% recall, a 98.4% Dice Coefficient, and a 96.9% Jaccard Index.

## 1. Introduction

Identification of genetically normal human blastocysts is a major current challenge to maximize the pregnancy rate of in vitro fertilization (IVF). Currently, embryo grading consists of embryologists manually evaluating the morphological features of blastocysts. This subjective and qualitative approach is prone to human bias and has only a limited correlation with the genetic quality of the embryo. An automated objective quality assessment process to benchmark embryos according to their potentials for either euploidy or live birth may help to boost IVF success rates and lower the risk of multiple pregnancies.

According to the traditional Gardner embryo grading system [1], the three most crucial blastocyst morphological parameters for predicting a successful pregnancy are: 1) the degree of blastocoel cavity expansion relative to the zona pellucida (ZP), 2) the compactness of the inner cell mass (ICM), and 3) the density of the trophectoderm epithelium (TE). To measure these parameters, image segmentation of the ZP, ICM, and TE of human blastocysts has been used. Santos et al. [2] demonstrated a level-set-based semi-automatic method for blastocyst component segmentation. Kheradmand et al. [3] used a two-layer neural network to segment ZP, TE, and ICM based on discrete cosine transform features. Saeedi et al. [4] presented automatic segmentation of the ICM and TE utilizing the texture information, and mathematical and statistical models. Singh et al. [5] demonstrated automatic segmentation of the TE using Retinex theory and a level-set algorithm. Moradi et al. [6] presented an automatic coarse-to-fine texture-based method to segment the ICM region. These works have improved the segmentation process, but the segmentation results can still be improved. Kheradmand et al. [7] focused on an ICM segmentation method based on a fully convolutional network pretrained by a 16-layer visual geometry group network [8]. This had improved segmentation results, but the overlap between segmented and manually labeled ICM was still limited to 76.5%. Moradi et al. [9] used a stacked dilated U-Net (SD U-Net) for ICM segmentation by incorporating a stack of 5 dilated convolution layers into the central bridge part of U-Net. An ensemble of SD U-Nets was also utilized to improve the final prediction map. This increased the overall performance of ICM segmentation compared to previous works [2, 3, 4, 6, 7], but there still is a 18.4% difference between automatic segmented and manually labeled ICM.

Although single-embryo transfer can provide similar pregnancy rates as double-embryo transfer without multiple pregnancies [10], it requires a TE biopsy and preimplantation genetic testing for aneuploidy (PGT-A). An alternative non-invasive approach is time-lapse imaging of the embryo [11]. Several time-lapse studies have reported parameters that correlate embryo viability [12], [13], [14]. Recently, Huang et al. [15] demonstrated an approach that utilizes early blastocyst expansion kinetics. They quantitatively analyzed the kinetics of blastocyst expansion in both euploid and aneuploid embryos from time-lapse images obtained before biopsy and PGT-A. Their results suggested that the analysis of expansion kinetics might improve the pregnancy rates after single embryo transfers either with or without biopsy. However, these assessments are labor intensive and would benefit from more objective, automated measurement of blastocyst expansion.

In work now described here, we developed an image-analysis tool that automates the assessment of blastocysts for the method introduced in [15]. The ZP of the analyzed blastocysts have been laser ablated prior to TE biopsy to check for chromosomal irregularities (Fig. 1). This provides quantitative information on the genetic quality of the blastocysts, but complicates measurement of the blastocyst

Md Yousuf Harun, M Arifur Rahman, Joshua Mellinger, Willy Chang, and Aaron T. Ohta are with the Department of Electrical Engineering, University of Hawaii at Manoa, Honolulu, HI, USA (e-mail: mdyousuf@hawaii.edu, aohta@hawaii.edu).

Thomas Huang, Brienne Walker, and Kristen Hori are with the Pacific IVF Institute, Kapiolani Medical Center, Honolulu, HI, USA (e-mail: huangt@hawaii.edu).

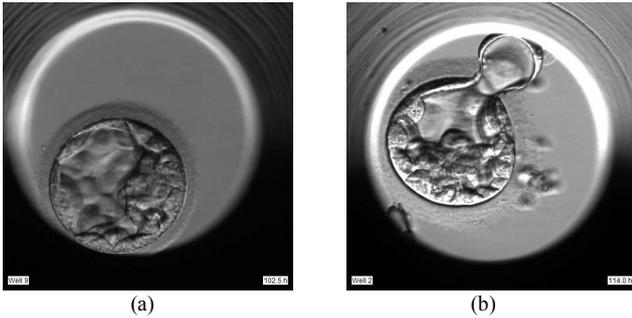

Figure 1. Images of (a) a zona-intact blastocyst and (b) a zona-ablated blastocyst.

area. The total area of blastocyst expansion is the sum of 1) the TE-enclosed area within the ZP, and 2) the TE-enclosed area herniating from the ablation slit. Non-biopsied embryos, which are circular in cross-section, can be accurately measured manually using the Embryoscope's (Vitrolife, USA) elliptical tool. However, this tool is less accurate at measuring the irregularly shaped TE-enclosed area herniating from the ablation slit. Moreover, the shape variability and unconstrained profile of blastocysts as well as the presence of fragments and artifacts surrounding them contribute to the complexities involved in this process. Thus, precise segmentation of zona-ablated blastocysts is a bottleneck in the morpho kinetic approach of embryo quality assessment and the IVF process.

To solve this problem, a deep-learning-based semantic segmentation approach was utilized, where each pixel of an image was labeled with a class of what it represents. A convolutional neural network (CNN) was trained to associate every pixel of an image with a class label. In this way, the CNN could precisely segment the irregularly shaped blastocyst. The specific CNNs used were various U-Net models, as U-Net has demonstrated significant success in medical image segmentation [16], [17]. The best four configurations of U-Net variants for this segmentation task were determined based on their performance on same test set. To the best of our knowledge, this is the first work on automated blastocyst segmentation of ZP-ablated blastocysts, as other reported works focused on segmentation of ZP, ICM, and TE regions in ZP-intact blastocysts.

## 2. SEGMENTATION METHOD

U-Net-based models were used for blastocyst segmentation. The pixels of blastocyst images were annotated according to two classes: blastocyst and background (non-blastocyst region). To maximize the network's performance on the dataset used here, its architecture and hyperparameters were optimized. To compare different U-Net variants, their basic architecture was configured similarly, and their hyperparameters were fine tuned.

### 2.1. Network Architecture

#### 2.1.1. U-Net

U-Net [18] is a widely used deep-learning architecture for biomedical image segmentation which can provide good results with a small number of training samples (Fig. 2). The encoder-decoder architecture of U-Net makes it robust for end-to-end segmentation. U-Net is composed of two symmetric contracting and expansive paths. The contracting path encodes the input image into a set of feature maps and the expansive path decodes the compact feature maps into a segmentation probability map.

#### 2.1.2. U-Net Variants

There are a lot of U-Net-based variants reported in the literature. For this segmentation task, U-Net, Dilated U-Net (SD U-Net), Residual U-Net (ResU-Net), and Residual Dilated U-Net (RD U-Net) were tested, as well as ensembles of the different U-Net models. The basic architecture of all U-Net variants is based on Ronneberger et al. [18]. The encoder contains four convolution blocks. Each block includes two convolution layers and a maxpooling layer. A similar convolution block is used to connect the encoding and decoding paths. The decoder has four upsampling convolution blocks. Each block includes one up-convolution layer followed by a concatenation with the corresponding encoded feature map from encoder and two convolution layers. Sigmoid activation is added to the final decoder layer. Batch normalization (BN), rectified linear unit (ReLu) activation, and dropout are applied to all other layers. The number of kernels in encoder-1 through encoder-4 were set to 16, 32, 64, and 128, respectively as shown in Fig. 2. Hence, the number of kernels in decoder-1 through decoder-4 are 128, 64, 32, and 16, respectively.

#### 2.1.3. Dilated U-Net

Dilated convolution [19] provides a wider receptive field and thereby, offers a wider access to the input. Moradi et al. [9] incorporated dilated convolution into the SD U-Net model. The central bridge part of the network includes a stack of five dilated convolution layers with dilation rates of 1, 2, 4, 8, and 16.

#### 2.1.4. Residual U-Net

Taking advantage of strengths from both deep residual learning [20] and the U-Net architecture, Zhang et al. [21] created a deep residual U-Net (ResU-Net) model. They used residual units instead of plain neural units to form a more robust architecture for better performance. The encoder, bridge, and decoder are built with residual units which comprises of two convolution blocks and a skip connection between input and output of the unit. Each convolution block has a BN layer, a ReLU activation layer and a convolutional layer.

#### 2.1.5. Dilated Residual U-Net

Similar as dilated U-Net, dilated convolution is applied to residual U-Net to increase the receptive field. Moradi et al. [22] presented a residual dilated U-Net that replaces the central bridge of U-Net by a stack of five dilated convolution layers with dilation rate of 1, 2, 4, 8, and 16.

#### 2.1.6. Ensemble of U-Net Variants

An ensemble of three U-Net variants: U-Net, ResU-Net, and RD U-Net was created using unweighted and weighted average schemes. Each model in the unweighted average ensemble contributed an equal amount to the final prediction map, which was generated by the arithmetic average of the base probability maps. In the weighted average ensemble, each ensemble member contributed to the final prediction in proportion to their estimated performances. The base probability map of each model is weighted based on their Jaccard Indices to construct the final probability map. Sometimes this approach improves the segmentation results.

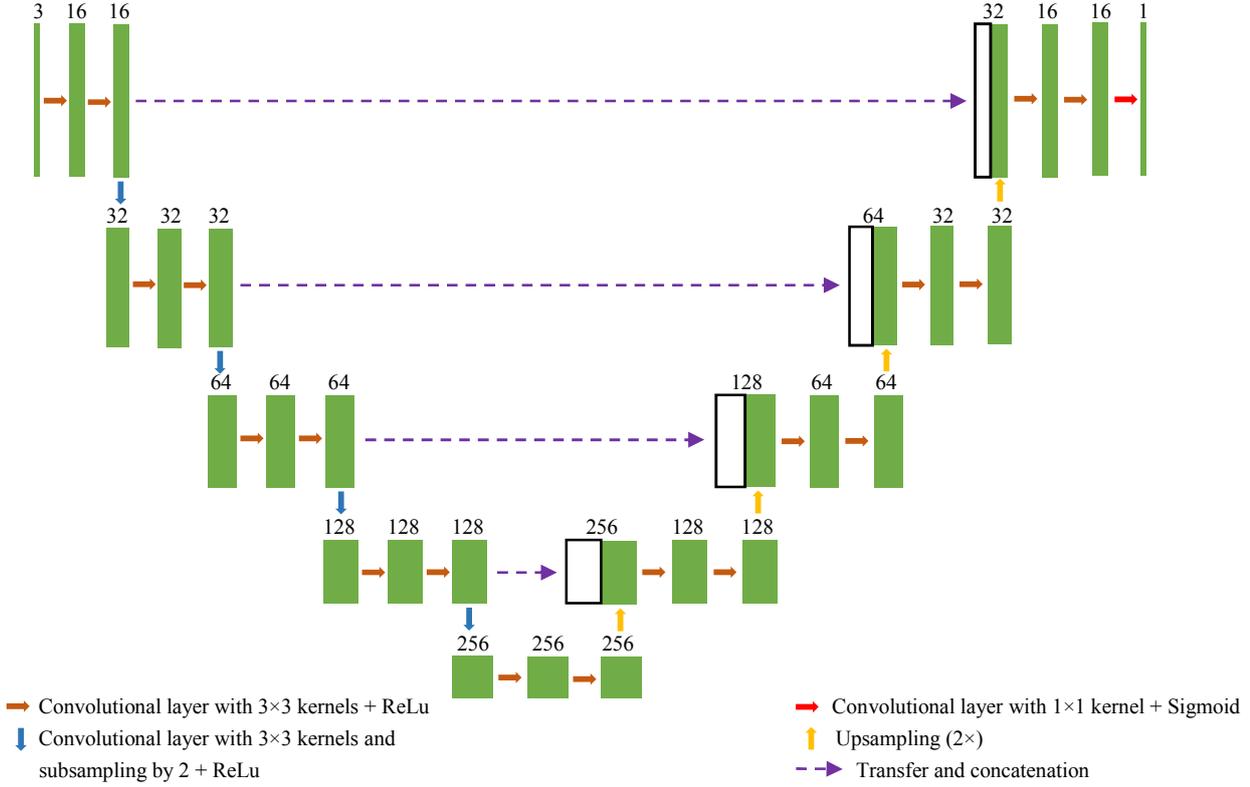

Figure 2. U-Net architecture.

## 2.2. Experimental Setup

### 2.2.1. Dataset and Ground Truth

The dataset used here consists of 617 embryoscope time-lapse images of 20 blastocysts, provided by the Pacific IVF Institute of the Kapiolani Medical Center. The embryos were cultured in LifeGlobal Medium (Guilford, USA) and monitored continuously over six days in an Embryoscope (Vitrolife, USA) using time-lapse imaging. The ZP of the embryos were focally ablated within the EmbryoSlide using a Lykos laser (Hamilton-Thorne, USA), when they reached blastocyst stage. The ZP-ablated blastocysts were then monitored via Embryoscope for the first 10.0 hours of expansion. The time-lapse images of each blastocyst during the 10.0 hours observation were finally converted to a video file for analyzing its expansion rate. After extracting frames from each recorded video file, there are 30 to 31 images of 500×500 resolution per blastocyst. The training set contains total 462 images (75% of the total dataset) and the test set consists of a total of 155 images (25% of the total dataset). The time-lapse images were annotated manually under the supervision of embryologists from the IVF Institute. The annotated images serve as the ground truth (GT) to evaluate the segmentation results of our method.

### 2.2.2. Framework

The models were trained on an NVIDIA GeForce GTX 1070 GPU with 8-GB memory and 16-GB RAM. The U-Net variants were implemented using Keras API with a Tensorflow backend. The networks were trained with a minibatch size of 16 and maximum epochs of 200.

## 2.3. Experiment

### 2.3.1. Data Preprocessing and Resizing

All the images were preprocessed with standard normalization before the training process. Then, the images were resized to 240×240 resolution to reduce the memory footprint on the GPU. Adjusting images according to network's receptive field enhances its performance. After segmentation, the images were restored to their original size (500×500).

### 2.3.2. Data Augmentation

To better handle the irregular cellular outlines in the ZP-ablated blastocysts, the networks are trained with additional artificially modified images. The images of the training set are transformed by using different augmentation techniques in Keras, including horizontal and vertical flip, rotation in a range up to 270 degrees, shifting horizontally and vertically up to 10% of width or height, and zooming up to 10% in size.

### 2.3.3. Image Shuffling

To improve network's generalization, the images in the training and test sets are shuffled before training begins. However, the distribution of images in the shuffled training and test sets is kept constant and the same test set is used to ensure a valid non-data dependent comparison between different CNNs. The training set images are also shuffled during each training epoch to improve the network's learning. It is worth noting that the input images and corresponding annotated ground truths maintained the same sequence in the training and test sets.

### 2.3.4. Hyperparameter Tuning

*1) Loss Function and Optimizer*

We adopt binary cross entropy Jaccard loss as loss function to reduce the loss between ground truth and network's prediction. The loss function helps to minimize the impact of class imbalance (blastocyst and background). An Adam optimizer [23] with an initial learning rate of 0.0001 is used to minimize the loss.

*2) Learning Rate Optimization*

We utilize the callback function integrated in Keras to reduce the learning rate by 5% once the learning stagnates and loss stops improving for maximum 5 epochs. The minimum learning rate is set to 0.000001.

*3) Dropout*

A 5% dropout is added to prevent overfitting.

*4) Early Stopping*

Early stopping is used to avoid over-fitting. The training is terminated when the network begins to overfit the data and loss stops decreasing for a continuous 15 epochs.

*5) Threshold Selection*

The threshold value was optimized over a range from 0.1 to 0.9 for the final probability map. The models are not sensitive to the choice of threshold, since the Jaccard index fluctuation is negligible for threshold values in a range from 0.4 to 0.6. Thus, the threshold was set to 0.5.

## 3. RESULTS

### 3.1. Performance Evaluation

The most widely used performance metrics in evaluating CNNs for semantic segmentation are: accuracy, precision, recall, Dice Coefficient, and Jaccard Index. Pixel accuracy reports the percent of pixels correctly classified as the background and blastocyst. Precision indicates the ratio of the blastocyst segmented by neural networks to a matching ground truth annotation. Recall describes how many of the blastocyst pixels annotated in ground truth are captured as positive predictions. The Dice Coefficient measures the similarity between predicted and labeled blastocyst regions. The Jaccard Index quantifies the percent overlap between predicted blastocyst and ground truth.

Hence, Dice Coefficient and Jaccard Index are very important measures for the overall quality of the segmentation results since they take into consideration the impact of both falsely identified and missed regions. Both metrics are equal to 1 if the segmented and ground truth regions completely overlap each other. In fact, they are correlated by Dice = (2 × Jaccard) / (Jaccard + 1). The performance metrics [24-27] are defined below:

$$Accuracy = \frac{TP + TN}{TP + TN + FP + FN} \quad (1)$$

$$Precision = \frac{TP}{TP + FP} \quad (2)$$

$$Recall = \frac{TP}{TP + FN} \quad (3)$$

$$Dice\ Coefficient = \frac{2 \times TP}{2 \times TP + FP + FN} \quad (4)$$

$$Jaccard\ Index = \frac{TP}{TP + FP + FN} \quad (5)$$

In equations (1-5), TP (true positives) represents the number of pixels correctly classified as blastocyst. TN (true negatives) indicates correctly extracted background pixels. FP (false positives) measures the number of pixels falsely identified as blastocyst. FN (false negatives) shows the missed blastocyst pixels.

### 3.2. Performance of U-Net Variants

The segmentation results on the test set are compared with the respective ground truths to evaluate the network's performance. Table 1 summarizes the performance of different U-Net models for blastocyst segmentation. While all models exhibited significant performance, RD U-Net achieved the best overall performance, with 98.8% recall and 96.9% Jaccard Index. It was found that the unweighted and weighted ensembles of U-Net variants do not provide better performance than RD U-Net.

### 3.3. Results of Blastocyst Segmentation

To better visualize the segmentation outcome, the results are organized into three categories: best (Jaccard Index of more than 97%), better (Jaccard Index from 95% to 97%), and fair (Jaccard Index from 90% to 95%) prediction, as shown in Table 2. Of the test set images, 51.6% fall into the best prediction category, 38.7% are in the better prediction category, and the remaining 9.7% are in the fair prediction category. The extracted blastocyst boundaries of ground truth are overlaid on the predicted blastocyst (by RD U-Net) to depict the discordance between them.

**Table 1.** Performance of different U-Net models evaluated on same test set.

| Model | Accuracy (%) | Precision (%) | Recall (%) | Dice Coefficient (%) | Jaccard Index (%) |
|---|---|---|---|---|---|
| U-Net | 99.3 | 98.0 | 98.2 | 98.1 | 96.3 |
| SD U-Net | 99.1 | 97.8 | 97.9 | 97.8 | 95.8 |
| ResU-Net | 99.4 | 98.1 | 98.7 | 98.4 | 96.8 |
| RD U-Net | **99.4** | **98.1** | **98.8** | **98.4** | **96.9** |
| Unweighted Ensemble | 99.2 | 97.8 | 98.3 | 98.1 | 96.2 |
| Weighted Ensemble | 99.2 | 97.8 | 98.3 | 98.1 | 96.2 |

**Table 2.** Segmentation results: Here, dark cyan, light green, yellow, and red highlight background, annotated blastocyst (ground truth), segmented blastocyst by RD U-Net, contour of ground truth, respectively. Jaccard Index (JI) and Dice Coefficient (DC) are mentioned in the performance categories since other performance metrics are significantly high.

| | Input Image | Ground Truth | Prediction | Comparison | |
|---|---|---|---|---|---|
| Best Prediction (Jaccard Index > 97%) | 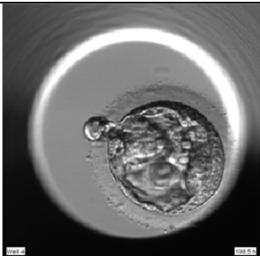 | 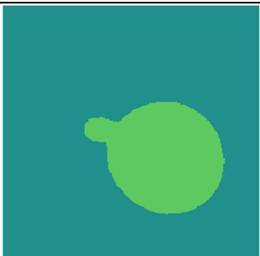 | 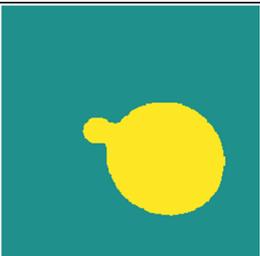 | 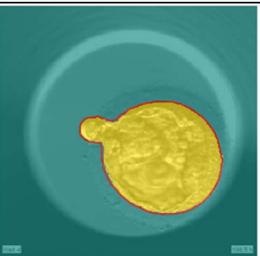 | JI 98.5% (highest), DC 99.3% |
| | 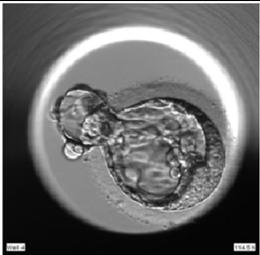 | 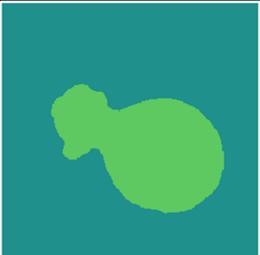 | 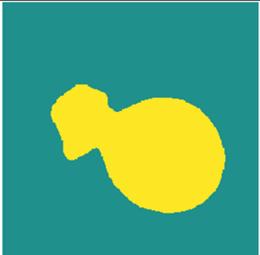 | 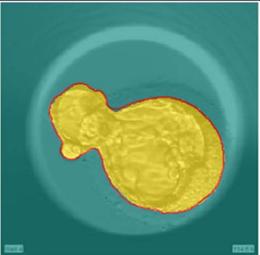 | JI 97.8%, DC 98.9% |
| Better Prediction (Jaccard Index > 95%) | 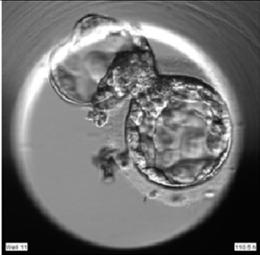 | 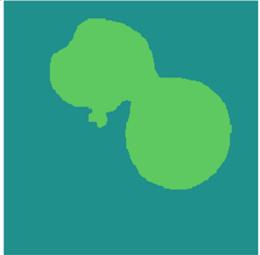 | 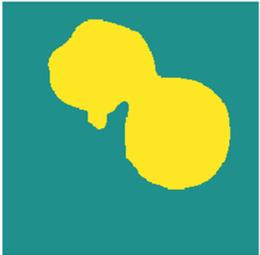 | 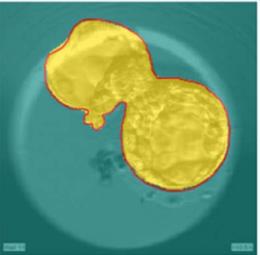 | JI 96.9%, DC 98.5% |
| | 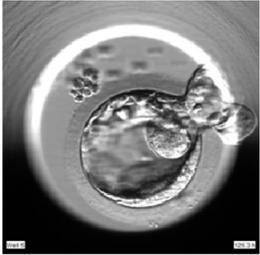 | 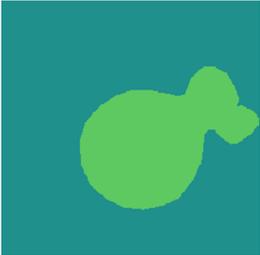 | 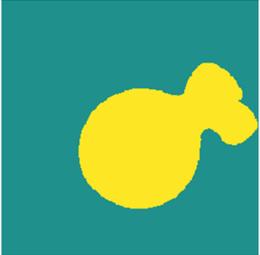 | 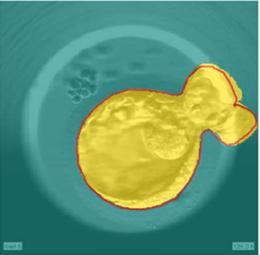 | JI 95.5%, DC 97.7% |
| Fair Prediction (Jaccard Index > 90%) | 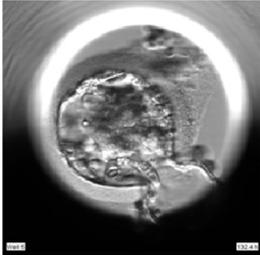 | 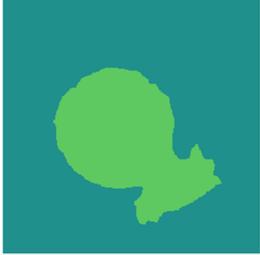 | 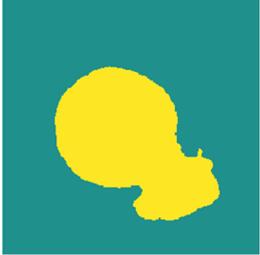 | 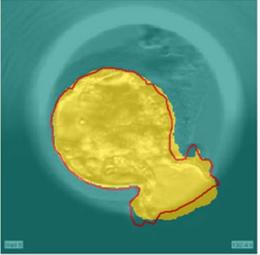 | JI 90.6% (lowest), DC 95.1% |
| | 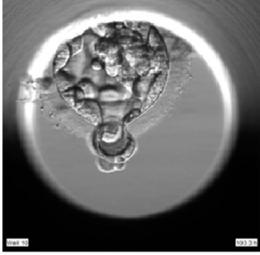 | 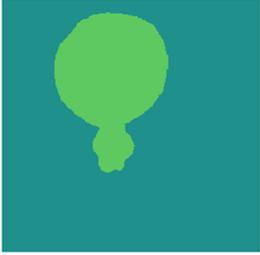 | 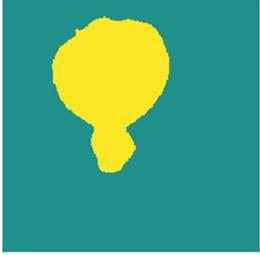 | 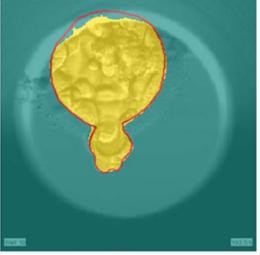 | JI 93%, DC 96.4% |

## 4. Conclusion

In this work, we presented an automated segmentation method to precisely identify the expanded blastocyst in the time-lapse images of zona-ablated blastocysts. Utilizing U-Net models, useful segmentation results were achieved on a relatively small dataset, with up to a 96.9% Jaccard Index and a 98.4% Dice Coefficient. This is the first work on automated image segmentation of ZP-ablated human blastocysts. This can provide vital morphological information to facilitate the automated analysis of blastocyst expansion kinetics for IVF embryo quality assessment.